\documentclass[11pt]{article}
\usepackage{amsmath,amssymb}
\usepackage{setspace,subfig}
\usepackage[hidelinks]{hyperref}
\usepackage{color}
\usepackage{booktabs}
\usepackage{pgf}
\usepackage{setspace}
\usepackage{tikz}
\usepackage{times}
\usepackage{url}
\usetikzlibrary{arrows,shapes.arrows,shapes.geometric,
	shapes.multipartb,backgrounds,decorations.pathmorphing,positioning,fit,automata}
\tikzset{
	>=stealth',
	true/.style={
		rectangle,
		draw=black, very thick,
		text width=6.5em,
		minimum height=2em,
		text centered,
		fill=gray, opacity = 0.5},
	punkt/.style={
		rectangle,
		rounded corners,
		draw=black, very thick,
		text width=6.5em,
		minimum height=2em,
		text centered},
	est/.style={
		circle,
		draw=black, very thick,
		text centered},
	shade/.style={
		circle,
		draw=black, very thick, fill=gray!50,
		text centered},
	weight/.style={
		circle,
		draw=black, very thick,
		text width=6.5em,
		minimum height=2em,
		text centered},
	pil/.style={
		->,
		thick,
		shorten <=2pt,
		shorten >=2pt,},
	double/.style={
		<->,
		thick,
		shorten <=2pt,
		shorten >=2pt,},
	dash/.style={
		dashed,
		thick,
		shorten <=2pt,
		shorten >=2pt,},
	dashdouble/.style={
		<->,
		dashed,
		thick,
		shorten <=2pt,
		shorten >=2pt,}
}

\usepackage{amsthm}
\usepackage{enumitem}
\usepackage{amssymb}
\usepackage{natbib}

\usepackage{caption} 

\usepackage{graphicx}
\newcommand{\ind}{\rotatebox[origin=c]{90}{$\models$}}

\newtheoremstyle{note}
{8pt}
{8pt}
{}
{}
{\bfseries}
{:}
{.5em}
{}

\theoremstyle{note}
\newtheorem{theorem}{Theorem}

\newtheorem{remark}{Remark}
\newtheorem{assumption}{Assumption}

\newtheorem{corollary}{Collorary}
\newtheorem{proposition}{Proposition}

\allowdisplaybreaks
\newcommand\numberthis{\addtocounter{equation}{1}\tag{\theequation}}

\usepackage[left=1in,right=1in,top=1in,bottom=1in]{geometry}
\usepackage{algorithm}
\date{}
\usepackage{tikz}

\definecolor{mygreen}{RGB}{144,241,47}

\newcommand{\tred}{\textcolor{red}}

\newcommand{\nind}{\not\!\perp\!\!\!\perp}
\newcommand{\Pn}{\mathbb{P}_n}

\usepackage[none]{hyphenat}

\usepackage[figuresright]{rotating}
\usepackage[utf8]{inputenc}
\usepackage{authblk}
\providecommand{\keywords}[1]
{
  \small	
  \textbf{\textit{Keywords---}} #1
}
\begin{document}

	\title{IV estimation of causal hazard ratio
}
	\author[1]{Linbo Wang} 
	\author[2]{Eric Tchetgen Tchetgen}
 	\author[3]{Torben Martinussen}
 	\author[4]{Stijn Vansteelandt}
	\affil[1]{Department of Statistical Sciences, University of Toronto, Toronto,
Ontario M5S 3G3, Canada}
	\affil[2]{Department of Statistics, University of Pennsylvania, Philadelphia, Pennsylvania 19104, U.S.A.}
	\affil[3]{Department of Biostatistics, University of Copenhagen, {\O}ster  Farimagsgade 5, 1014
Copenhagen, Denmark}
	\affil[4]{Department of Applied Mathematics, Computer Science and Statistics, Ghent University,
Krijgslaan 281 (S9),\\ 9000 Ghent, Belgium}
	\clearpage \maketitle


\begin{abstract}
Cox's proportional hazards model is one of the most popular statistical models to evaluate associations of exposure with a censored failure time outcome. When confounding factors are not fully observed, the exposure hazard ratio estimated using a Cox model is subject to unmeasured confounding bias. To address this, we propose a novel approach for the identification and estimation of the causal hazard ratio in the presence of unmeasured confounding factors. Our approach is based on a binary instrumental variable, and an additional no-interaction assumption in a first stage regression of the treatment on the IV and unmeasured confounders. We propose, to the best of our knowledge, the first consistent estimator of the (population)   causal hazard ratio within an instrumental variable framework. A version of our estimator admits a closed-form representation. We derive the asymptotic distribution of our estimator, and provide a consistent estimator for its asymptotic variance.  Our approach is illustrated via simulation studies and a data application.
\end{abstract}

\keywords{
Causal inference; Cox model;  Marginal structural model;  Survival analysis; Unmeasured confounding}

\maketitle

\section{Introduction}

In observational studies with a possibly right-censored outcome, the Cox proportional hazards model is by far the dominant analysis tool to infer the association between a treatment and an outcome. The associational measure here is the hazard ratio, that is, the ratio of instantaneous incidence rates between treatment groups. It is well-known that in observational studies, the hazard ratio estimated with a Cox model may be subject to unmeasured confounding. 

A classical approach to deal with unmeasured confounding uses an instrumental variable. Intuitively, conditional on baseline covariates, an instrumental variable is an exogenous variable that is associated with the outcome only through its association with the treatment. The instrumental variable approach has been well-developed for the analysis of continuous and binary outcomes \citep[e.g.][]{wright1928tariff,angrist1996identification,abadie2003semiparametric,hernan2006instruments,wooldridge2010econometric,wang2018bounded} but less so for a right-censored survival outcome, particularly within the dominant Cox regression framework. 
This is mainly because the commonly used two-stage methods for instrumental variable estimation fail to provide consistent estimates  due to non-collapsibility of the hazard ratio.

In this paper, we fill this gap by proposing a consistent estimator of the population-average causal hazard ratio in the case of an endogenous treatment variable, which to the best of our knowledge, is the first in the literature.
We make the proportional causal hazard ratio assumption, which results in the so-called marginal structural Cox model \citep{hernan2000marginal}. 
The marginal structural Cox model parameter can be interpreted as the causal hazard ratio.
To identify the causal hazard ratio with a binary endogenous treatment variable, in addition to a valid binary instrument, we require a no-interaction assumption that the instrument and unmeasured confounders do not interact on the additive scale for their effects on the exposure. Our identification result builds on that of \cite{wang2018bounded}, who establish identifiability of the average treatment effect on the additive scale under a similar assumption. Our assumption allows the outcome model to be completely unrestricted other than the marginal structural Cox model assumption, thus in sharp contrast to various treatment effect homogeneity assumptions previously used in the literature to identify population-average treatment effects with an instrument \citep[e.g.][]{aronow2013beyond,hernan2006instruments}.  Our identification formula readily leads to an estimating equation for the causal hazard ratio. To ease computation,  
we also develop a closed-form representation of the causal hazard ratio under our identification assumption. Although this might not directly improve the efficiency of the resulting estimator, it is particularly appealing computationally as without a closed-form representation, in practice it can be difficult to find a solution to an estimating equation. Even if one finds one solution, it can be difficult to check the uniqueness of such a solution.  Our results may also be extended in a number of important directions. For example, it can be used to identify the cumulative baseline hazard function, the causal hazard ratio conditional on baseline covariates, and the cause-specific causal hazard ratio in a competing risk setting. 

Our target of inference is different from the targets of most previous developments for instrumental variable estimation in a survival context, which are often motivated by randomized trials with non-compliance.  The treatment effects considered by these proposals are defined within the so-called complier stratum, consisting of individuals who would comply with the assigned treatment under both active treatment and control. Such estimands  include  the complier hazard difference \citep{baker1998analysis}, the complier hazard ratio \citep{loeys2003causal,cuzick2007estimating}, the complier quantile causal effect \citep{frandsen2015treatment,yu2015semiparametric}, the complier survival probabilities \citep{nie2011inference,yu2015semiparametric}  and the complier average causal effect \citep{abadie2003semiparametric,cheng2009semiparametric,yu2015semiparametric}.  
However, in practice, the complier causal effects are often only of secondary interest as they concern a highly selective {unknown} subset of the population \citep{robins1996identification}. Furthermore,  its definition depends on the particular instrument that is available \citep{wooldridge2010econometric}. This could potentially be a serious limitation outside of the non-compliance setting, especially when there is no natural choice of the instrument such as a randomized treatment assignment.  

Our work instead contributes to the literature on  instrumental variable estimation of population-average treatment effects in a survival context. Prior to our work,  \cite{robins1991correcting} parameterize the treatment effect under a structural accelerated failure time model,  \cite{li2015instrumental}, \cite{tchetgen2015instrumental} and \cite{martinussen2017instrumentalcumulative} consider estimating the conditional hazard difference under a structural cumulative survival model, \cite{martinussen2019instrumentalcox}, \cite{sorensen2019causal} and \cite{martinez2019adjusting} consider estimating the causal hazard ratio among the treated, while \cite{choi2017estimating} consider estimating the average treatment effect on the survival time. None of these methods, however, were designed to estimate the population-average causal hazard ratio, which is a natural target of inference given the popularity of the Cox model in practice. 
Although \cite{mackenzie2014using} have also considered instrumental variable estimation of the population-average causal hazard ratio, their estimating equation is only approximately unbiased under certain conditions.  Furthermore, their approach relies on an instrument valid for estimating the effects of all the covariates, and is limited to a somewhat artificial causal model \citep{tchetgen2015instrumental}.

\section{Background}
\label{sec:framework}

\subsection{Framework and notation}

Consider an observational study where the interest lies in estimating the effect of a binary treatment $D$ on a possibly censored continuous survival outcome  $T$. The effect of interest is subject to confounding by (a subset of) observed variables $X$ as well as unobserved variables $U$. Let $C$ denote the censoring time and $\Delta$ be the event indicator: $\Delta = I(T\leq C).$ The observed time    $Y = \min(T,C).$ Let $Z$ denote a binary instrumental variable with a 0-1 coding scheme.
Using the framework of the potential outcome, let $D(z)$ be the potential exposure if the instrument had taken value $z$ to be well-defined \citep[the Stable Unit Treatment Value Assumption,][]{rubin1980comment}. Similarly, we assume $T(d)$ and $C(d)$, the potential survival and censoring time if a unit were exposed to $d$ to be well-defined. 
The potential survival function is defined as $S^T_d(t) = P(T(d) \geq t)$,  and the potential hazard function is defined  as $\lambda^T_d(t) = -[S^T_d(t)]^\prime / S^T_d(t).$  Let $Y(d) = \min \{T(d), C(d)\}$. We may then similarly define $S_d^Y(y)$ and $\lambda_d^Y(y)$.

We assume the marginal structural Cox model:
\begin{equation}
\label{eqn:mscm}
\lambda^T_d(t) = \lambda^T_0(t) e^{\psi d}.
\end{equation}
We are interested in estimating $\psi$,  the log of the causal hazard ratio. 

We make the following assumptions commonly invoked in an instrumental variable analysis. 
\begin{assumption}[Independence]
	\label{assumption:independence}
	$Z \ind U \mid X.$
\end{assumption}
\begin{assumption}[Instrumental relevance]
	\label{assumption:relevance}
	$Z \nind D\mid X=x,$ for all $x$ in the support of $X$.
\end{assumption}
\begin{assumption}[Sufficiency of $U$]
	\label{assumption:U}
	$T(d), C(d)\ind (D,Z)\mid (X,U).$
\end{assumption}
\begin{assumption}[Independent censoring]
	\label{assumption:ind_censoring}
	$
	C(d) \ind T(d), d=0,1.
	$
\end{assumption}
We note that under the consistency assumption,  Assumption \ref{assumption:U} implies the exclusion restriction assumption $Z \ind (T,C) \mid D,U,X.$ 
Figure \ref{DAG:iv_model} gives a simple illustration  of the
conditional instrumental variable model. Under the consistency assumption, Assumptions \ref{assumption:independence}--\ref{assumption:ind_censoring} can be read off from the single world intervention graph \citep{richardson2013single} in Figure \ref{DAG:iv_model}(b) via d-separation \citep{pearl2009causality}. As pointed out by a reviewer, the causal graphs in Figure \ref{DAG:iv_model} need not be faithful; in particular, none of the links $X\rightarrow Z,  X\rightarrow D$ or $X\rightarrow T$ is necessary. 
 See Figure \ref{DAG:iv_model2} in the Supplementary Material for an alternative causal graphical model that satisfies Assumptions \ref{assumption:independence}--\ref{assumption:ind_censoring}.

\begin{figure}[!htbp]
	\parbox{.4\textwidth}{
		\vspace{0.9cm}
		\centering
		\begin{tikzpicture}[->,>=stealth',node distance=1cm, auto,]
		\node[est] (Z) {$Z$};
		\node[est, right = of Z] (D) {$D$};
		\node[est, right = of D] (T) {$T$};
		\node[shade, below = of D] (U) {$U$};
		\node[est, above = of D] (V) {$X$};
		\node[est, above = of T] (C) {$C$};
		\path[pil] (Z) edgenode {} (D);
		\path[pil] (D) edgenode {} (T);
		\path[pil] (U) edgenode {} (D);
		\path[pil] (U) edgenode {} (T);
		\path[pil] (V) edgenode {} (Z);
		\path[pil] (V) edgenode {} (D);
		\path[pil] (V) edgenode {} (T);
		\path[pil] (D) edgenode{} (C);
		\path[double] (Z) edge [bend right] node  {} (D);
		\end{tikzpicture}
		\quad \\ \bigskip (a). A directed  acyclic graph with a bi-directed arrow.
	}
	\parbox{.4\textwidth}{
		\centering	
		\begin{tikzpicture}[>=stealth, ultra thick, node distance=2cm,
		pre/.style={->,>=stealth,ultra thick,black,line width = 1.5pt}]
		\begin{scope}
		\node[name=Z,shape=ellipse splitb, ellipse splitb/colorleft=black, ellipse splitb/colorright=red, 
		ellipse splitb/innerlinewidthright = 0pt,  
		/tikz/ellipse splitb/linewidthright = 1pt,   
		ellipse splitb/gap=3pt, style={draw},rotate=90] {
			{\rotatebox{-90}{$Z$\;}}
			\nodepart{lower}
			{\rotatebox{-90}{\;\textcolor{red}{$z$}}}
		};
		\node[name=D,shape=ellipse splitb, below right=0.65cm and 1.8cm of Z,  ellipse splitb/colorleft=black, ellipse splitb/colorright=red, 
		ellipse splitb/innerlinewidthright = 0pt,  
		/tikz/ellipse splitb/linewidthright = 1pt,   
		ellipse splitb/gap=3pt, style={draw},rotate=90] {
			{\rotatebox{-90}{$D(\tred{z})$\;}}
			\nodepart{lower}
			{\rotatebox{-90}{\;\textcolor{red}{$d$}}}
		};
		\node[thick, name=T,shape=ellipse,style={draw}, below right =0.05cm and 1.8cm  of D,outer sep=0pt, text width = 8mm]
		{$T({\color{red}{d}})$
		};
		\node[thick, name=C,shape=ellipse,style={draw}, above  =1.2cm  of T, outer sep=0pt, text width = 8mm]
		{$C({\color{red}{d}})$
		};
		\draw[pil,->] (Z) to (D);
		\node[est, above right=1.5cm and 0.3cm of D] (V) {$X$};
		\path[pil] (Z) edgenode {} (D);
		\path[pil] (D) edgenode {} (T);
		\path[pil] (D) edgenode {} (C);
		\draw[pil, ->] (2.7,2)  to[bend right]  (-0.5,0.4);
		\draw[pil, ->] (3.1,1.7)  --  (3.1,0.4);
		\draw[pil, <->] (3.1,-0.5)  to[bend left]  (-0.5,-0.4);
		\path[pil] (V) edgenode {} (T);
		\node[shade, below = of D] (U) {$U$};
		\path[pil] (U) edgenode {} (3.3,-0.5);
		\path[pil] (U) edgenode {} (T);
		\end{scope}
		\end{tikzpicture}
		\quad \\ \bigskip	(b). A single world intervention graph with a bi-directed arrow.
	}
	\caption{Causal graphs representing the instrumental variable model defined by Assumptions \ref{assumption:independence} -- \ref{assumption:ind_censoring}. The bi-directed arrow between $Z$ and $D$ indicates potential unmeasured common causes of $Z$ and $D$. Variables $X,Z,D$ are observed; $T$ is possibly right censored; $U$ is unobserved. The left panel gives a causal directed acyclic graph \citep{pearl2009causality} with a bi-directed arrow, and the right panel gives a single world intervention graph \citep{richardson2013single} with a bi-directed arrow. This figure appears in color in the electronic version of this article,
and any mention of color refers to that version.}
	\label{DAG:iv_model}
\end{figure}

One can see from the bi-directed arrows in Figure  \ref{DAG:iv_model} that we allow for latent common causes of $Z$ and $D$, so that the instrument $Z$ and exposure $D$ can be associated because $Z$ has a causal effect on $D$, {or} because they share a common cause, or both. { This is important as in observational study settings,
	it	may not be realistic to assume that one has measured all common causes of $Z$ and $D$.}  

To focus on the main challenges introduced by unmeasured confounding, we have assumed independent censoring as in Assumption \ref{assumption:ind_censoring}. It can be extended to allow for censoring dependent on observed covariates $X$, that is,
\begin{equation}
    \label{eqn:ignorable-censoring}
    C(d) \ind T(d)\mid X, d=0,1.
\end{equation}  See Proposition \ref{cor:ignorable-censoring} for details.

Even with a valid instrument, in general, population-level causal effects are not identifiable from observed data. We now review some existing methods for identifying the average treatment effect in the literature.

\subsection{Instrumental variable methods}

With a continuous outcome $Y$, a classical method to estimate the population average treatment is based on the following system of linear structural equation models:
 \begin{subequations}
\begin{flalign}
    D &= \alpha_0 + \alpha_1 Z+\alpha_2 X + \alpha_3 U + \epsilon_D; \label{eqn:lsem1} \\
    Y &= \beta_0 + \beta_1 D + \beta_2 X + \beta_3 U + \epsilon_Y. \label{eqn:lsem2}
\end{flalign}
\end{subequations}

The two-stage least squares (TSLS) method then proceeds as follows: one first regresses the treatment $D$ on the instrument $Z$ and covariates $X$ to obtain $\widehat{D},$ and then regresses $Y$ on $\widehat{D}$ and $X$ to get the treatment effect estimate $\widehat{\beta}_1.$  To illustrate the idea behind the TSLS, consider the simple case without  $X$ and $\epsilon_D$. Without loss of generality, assume $E(U) = 0$ so that $\widehat{D}$ is approximately $\alpha_0+\alpha_1 Z.$ In this case, equations \eqref{eqn:lsem1} and \eqref{eqn:lsem2} imply that 
$
    Y = \beta_0 + \beta_1 \widehat{D} + (\alpha_3\beta_1 + \beta_3) U + \epsilon_Y. 
$
Since $Z\ind (U, \epsilon_Y),$ we have $\widehat{D} \ind (U,\epsilon_Y)$ and   $E[Y\mid \widehat{D}] = \beta_0 + \beta_1 \widehat{D}.$ Hence regressing $Y$ on $\widehat{D}$ yields a consistent estimate for $\beta_1.$ 

The TSLS method, however, cannot be directly extended to obtain a consistent estimate for the causal hazard ratio. To see this, again   consider the simple case without $X$ and $\epsilon_D$. Instead of \eqref{eqn:lsem2}, one may assume a structural Cox model:
\begin{equation}
    \label{eqn:cox2}
    \log\lambda(y\mid D, U, Z) = \log\lambda_0(y) + \gamma_1 D + \gamma_2 U.
\end{equation}
Although under \eqref{eqn:lsem1}, we have $\log\lambda(y\mid \widehat{D}, U) = \log\lambda_0(y) + \gamma_1 \widehat{D} + (\alpha_3\gamma_1 + \gamma_2) U,$ due to non-collapsibility of hazard ratio, in general $\log\lambda(y\mid \widehat{D}) \neq  \log\lambda_0(y) + \gamma_1 \widehat{D}$. In this case, a two-stage predictor substitution (TSPS) method that fits a Cox model of $Y$ on $\widehat{D}$ yields a biased estimate of $\gamma_1$ unless under degenerate circumstances such as $\gamma_1 = 0$ or $\alpha_3\gamma_1+\gamma_2 = 0$. We refer interested readers to \citet[][\S 4.3]{wan2018general} for a detailed discussion of the bias from the TSPS method in the general case.

\begin{remark}
    We further note that in the simple case without $X$ and $\epsilon_D,$ $\gamma_1$ in \eqref{eqn:cox2} denotes the log of the causal hazard ratio conditional on $D,U,Z$. Due to   non-collapsibility of the hazard ratio, it is generally different from the log of the marginal causal hazard ratio, $\psi$.
\end{remark}

In the linear case,
the structural equation models \eqref{eqn:lsem1} and \eqref{eqn:lsem2} imply the following no-interaction assumptions:
 \begin{subequations}
\begin{flalign}
    E[D\mid Z=1, X, U] -E[D\mid Z=0, X, U] \ind U; \label{eqn:ne1} \\
    E[Y \mid D=1,X, U] -E[Y\mid D=0, X, U] \ind U. \label{eqn:ne2}
\end{flalign}
\end{subequations}
\cite{wang2018bounded} show that  with a binary treatment $D$, if either \eqref{eqn:ne1} or \eqref{eqn:ne2} holds, the average treatment effect $\beta_1$ can be identified and satisfies
\begin{equation}
\label{eqn:weighted}
    \beta_1 = E \left\{ \dfrac{2Z-1}{f(Z\mid X) \delta^D(X)} Y \right\},
\end{equation}
where $\delta^D(X) = E[D\mid Z=1,X] - E[D\mid Z=0, X]$ and $f(Z\mid X)$ denotes the conditional density of $Z$ given $X$. We shall now build on \cite{wang2018bounded}'s work to identify the causal hazard ratio.

\section{Identification and estimation of the causal hazard ratio}
\label{sec:identification}

\subsection{Identification of the causal hazard ratio}
\label{subsec:e.e}

We now consider the identification problem of the log causal hazard ratio $\psi.$ 
The results of \cite{wang2018bounded} imply that the average causal effect is identifiable without imposing assumptions on the outcome model, thus circumventing the difficulty introduced by non-collapsibility of the hazard ratio. Motivated by this, we shall pursue identification of the log causal hazard ratio $\psi$ under the assumption \eqref{eqn:ne1} on the treatment generating model, which we formally state below.
\begin{assumption}
	\label{assumption:no-interaction}
For $U$ that satisfies Assumptions \ref{assumption:independence},\ref{assumption:U}, there is no additive $U-Z$  interaction in $E[D\mid Z, X,U]$:
	\begin{equation}
	\label{eqn:iden}E[D\mid Z=1,X,U] - E[D\mid Z=0,X,U]  = E[D\mid Z=1,X] - E[D\mid Z=0, X].
	\end{equation}
\end{assumption}

	When the instrument $Z$  is randomized, \eqref{eqn:iden} is equivalent to $E[D(1)-D(0)\mid X,U] = E[D(1)-D(0)\mid X]$. Let $(D(1), D(0))$ be the compliance type \citep[Table 1]{wang2018bounded}. Then Assumption \ref{assumption:no-interaction} holds as long as there are no unmeasured confounders that also predict compliance type.  The latter condition is closely related to the principal stratum homogeneity assumption that the causal effect is equal across principal strata \citep[e.g.][]{aronow2013beyond}, which holds if the compliance type itself is not an (unmeasured) confounder. Note that due to non-collapsibility of the hazard ratio, even under the principal stratum homogeneity assumption, the complier hazard ratio is generally not equal to the marginal hazard ratio. So methods that identify complier hazard ratio may not directly be used to identify the marginal causal hazard ratio.



\begin{remark}
	In general, there may be more than one set of unmeasured covariates $U$ that satisfy assumptions  \ref{assumption:independence} and \ref{assumption:U}. We say Assumption \ref{assumption:no-interaction} holds if at least one of these sets of covariates $U$ also satisfies \eqref{eqn:iden}.
\end{remark}

To identify the log hazard ratio $\psi$ under Assumption \ref{assumption:no-interaction}, recall that the partial score equation  in a regular Cox model \citep{cox1972regression} takes the following form: 
\begin{equation}
\label{ee:cox}   		H(\tau) = \mathbb{P}_n \int  \left[  W -  \dfrac{ \mathbb{P}_n \left\{ W e^{\tau W} I(Y \geq y)  \right\}} { \mathbb{P}_n\left\{ e^{\tau W} I(Y \geq y) \right\} } \right]  dN(y),
\end{equation}
where $N(y) = I(Y\leq y, \Delta=1)$ is the counting process of observed failure events, $\mathbb{P}_n$ denotes empirical average and $W$ denotes the covariates in a regular Cox model. 

Motivated by \eqref{eqn:weighted},  we consider a weighted version of \eqref{ee:cox} by applying some weight function $\omega(Z,X,D)$ to the at risk process $I(Y\geq y)$ for each time point $y$. This results in the following estimating equation:
	\begin{equation}
	\label{ee:eric}
	H\left( \psi \right) =\mathbb{P}_n\int dN(y) \omega(Z,X,D)\left[ D-\dfrac{\mathbb{P}_n\left\{ De^{\psi D}  I(Y\geq y)\omega(Z,X,D) \right\} }{\mathbb{P}_n\left\{ e^{\psi D} I(Y\geq y)\omega(Z,X,D) \right\} }%
	\right].
	\end{equation}
A natural choice for the weight function would be to use $\omega_0(Z,X) = (2Z-1)/\{f\left( Z|X\right) \delta^D \left( X\right) \}$, as in \eqref{eqn:weighted}. However, it will make \eqref{ee:eric} ill-defined under large samples since  $E\left\{ e^{\psi D} I(Y\geq y)\omega_0(Z,X) \right\}=0$ under the null that $\psi=0$.  To solve this problem, we add a stabilization term ${h}(D)$  to the weight function to ensure that \eqref{ee:eric} is well-defined under large samples. Theorem  \ref{thm:identification} shows that the log causal hazard ratio $\psi$ can indeed be uniquely  identified via the population version of the estimating equation \eqref{ee:eric}.

\begin{theorem}
	\label{thm:identification}
	Under the marginal structural Cox model \eqref{eqn:mscm} and Assumptions \ref{assumption:independence}--\ref{assumption:no-interaction}, the causal hazard ratio is identifiable  and  is the unique solution to  $E\{H(\psi)\} = 0,$ where $H(\psi)$ is defined in \eqref{ee:eric},
$\omega(Z,X,D) = {h}(D)(2Z-1)/\{f\left( Z|X\right) \delta^D \left( X\right) \}$ and ${h}(D)$ is any  function of $D$ such that ${h}(1){h}(0)<0$.
\end{theorem}

 Our weighted analyses here and in \eqref{eqn:weighted} are similar in spirit to inverse probability weighting techniques commonly used in survival analysis to account for censoring \citep{robins1992recovery}, observed confounding \citep{hernan2000marginal} and to detect early differences in survival times \citep[weighted log-rank test, e.g.][]{fleming2011counting}.

Identification formula \eqref{ee:eric} directly  leads to a weighting estimator for $\psi$. Suppose $f(Z\mid X;\eta)$ and $\delta^D(X; \beta)$ are finite-dimensional models on $f(Z\mid X)$ and $\delta^D(X),$ respectively. The parameter $\eta$ can be estimated using the maximum likelihood estimator $\widehat{\eta}.$ The conditional risk difference model $\delta^D(X; \beta)$, however, does not give rise to a likelihood by itself, so estimation of $\beta$ relies on additional nuisance models. Choosing an appropriate nuisance model for estimating $\delta^D(X;\beta)$ is non-trivial,  as a naive nuisance model on $p_0^D(X) = P(D=1\mid Z=0,X)$  is not desirable: given models on $p_0^D(X)$ and $\delta^D(X),$ there is no guarantee that $p_1^D(X) = P(D=1\mid Z=1, X) = p_0^D(X) + \delta^D(X)$ lies in the unit interval $[0,1].$ Instead, \cite{richardson2017modeling} develop a  nuisance model $OP^D(X;\zeta),$ where $OP^D(X) = p_1^D(X) p_0^D(X) / \{(1-p_1^D(X))(1-p_0^D(X))\}.$ 
It can be shown that with this parameterization, $p_1^D(X;\beta, \zeta)$ is guaranteed to lie in the unit interval, so that the MLE $(\widehat{\beta},\widehat{\zeta})$ may be obtained by unconstrained maximization. 
Alternatively, one may model $P(D=1\mid Z,X)$ directly using say, a logistic regression and then obtain a plug-in estimate for $\delta^D(X).$

\begin{remark}
    Although it seems more straightforward to use logistic regression models on $P(D=1\mid Z,X)$ to estimate $\delta^D(X),$  later in the simulations, we simulate data by specifying  models on  $\delta^D(X,U)$ and $OP^D(X,U)$ as it is easier to impose Assumption \ref{assumption:no-interaction} this way. In particular, we simply let $\delta^D(X,U) = \delta^D(X)$ while $P(D=1\mid Z,X,U)$ still depends on $U$ via $OP^D(X,U).$ If instead, one simulates data by specifying logistic models on $P(D=1\mid Z,X,U)$ directly, then to impose Assumption \ref{assumption:no-interaction}, one would typically need to assume that $P(D=1\mid Z=z,X,U)$ is independent of $U$, in which case $U$ is not a confounder.
\end{remark}

Equation \eqref{ee:eric} motivates an  inverse probability weighting estimator, defined as a solution to the following equation:
\begin{equation}
\label{ee:eric_est}
\sum\limits_{i=1}^n {\Delta_i \widehat{\omega}(Z_i,X_i,D_i)}  
\left[ D_i-\dfrac{\sum\limits_{j=1}^n \left\{ D_je^{\psi D_j}  I(Y_j\geq Y_i) \widehat{\omega}(Z_j,X_j,D_j) \right\} }{\sum\limits_{j=1}^n \left\{ e^{\psi D_j} I(Y_j\geq Y_i) \widehat{\omega}(Z_j,X_j,D_j) \right\} }
\right]  = 0,
\end{equation}
where $\widehat{\omega}(Z_i,X_i,D_i) = {h}(D_i)(2Z_i-1)/\left\{f\left(Z_i|X_i; \widehat{\eta}\right) \delta^D \left( X_i; \widehat{\beta}\right) \right\}.$
Under suitable regularity conditions including correct specification of the models $f(Z\mid X;\eta), \delta^D(X;\beta), OP(X; \zeta)$, one can show that the solution to \eqref{ee:eric_est} is asymptotically linear using standard empirical process theory. 
In practice, however, it may be computationally cumbersome to solve equation \eqref{ee:eric_est}.  We address this problem in the next subsection by proposing an alternative estimator that is available in closed form.

\subsection{A closed-form representation of the causal hazard ratio}
\label{subsec:closed-form}


To simplify \eqref{ee:eric}, a natural idea is to  search for $h(D)$ such that 
\begin{equation}
    \label{eqn:goal}
    E\left\{ De^{\psi D}  I(Y\geq y)\omega(Z,X,D) \right\} =    E\left\{ \widetilde{g}(D)   I(Y\geq y)\omega_0(Z,X) \right\} = 0,
\end{equation}
where $\widetilde{g}(D) = h(D) D e^{\psi D}.$ 
If we can find such a $h(D)$, then \eqref{ee:eric} becomes
    $${H}(\psi) =\int dN(y)\omega_0(Z,X) \widetilde{g}(D) e^{-\psi D},$$
and $E\left\{{H}(\psi)\right\} = 0$ has a closed-form representation:
\begin{equation}
    \label{eqn:not}
     \exp(\psi) =  \dfrac{E\int dN(y) (-D)\omega_0(Z,X) \widetilde{g}(1)}{E\int dN(y) (1-D)\omega_0(Z,X) \widetilde{g}(0)}.
\end{equation}
Note, however, that it is not possible to identify $\exp(\psi)$ from equation \eqref{eqn:not}. This is because by construction, $\widetilde{g}(0) = 0$, so that the right hand side of equation \eqref{eqn:not} is not well-defined.


To solve this problem, instead of looking for $h(D),$ we shall directly look for a measurable function $g(D)$ so that $g(1)g(0)\neq 0$ and \eqref{eqn:goal} holds replacing  $\widetilde{g}(D)$ with $g(D).$ In other words, we look for $g(D)$ that is orthogonal to $I(Y\geq y) \omega_0(Z,X)$ in the space $L_2(Z,X,Y,D)$. In general, all such functions may be represented as 
$\{m(D)E\left\{I(Y\geq y)\omega_0(Z,X)\right\} -  E\left\{m(D) I(Y\geq y)\omega_0(Z,X)  \right\} : m(D) \text{ is measurable}  \}.$
Theorem \ref{thm:identification_torben} shows that as long as $m(1) \neq m(0)$ so that $g(1)g(0) \neq 0,$ \eqref{eqn:not} holds replacing  $\widetilde{g}(D)$ with $g(D).$

\begin{theorem}
	\label{thm:identification_torben}
	Under the marginal structural Cox model \eqref{eqn:mscm} and Assumptions \ref{assumption:independence}--\ref{assumption:no-interaction}, we have
	\begin{equation}
	\label{eqn:identification_torben}
	\exp(\psi) = \dfrac{ E \int dN(y) (-D) \omega_0(Z,X) \left\{m(1) \gamma_1(y) - \gamma_2^m(y) \right\} }
	{E \int dN(y) (1-D) \omega_0(Z,X) \left\{m(0) \gamma_1(y) - \gamma_2^m(y) \right\} },
	\end{equation}
	where $\gamma_1(y) = E\left\{I(Y\geq y)\omega_0(Z,X)\right\}, \gamma_2^m(y) = E\left\{m(D) I(Y\geq y)\omega_0(Z,X)  \right\}$ and  $m(1)\neq m(0)$.
\end{theorem}

Theorem \ref{thm:identification_torben} can be extended in several directions.
First, it can be extended to identify the cumulative baseline hazard function.
\begin{proposition}
\label{proposition:breslow}
	Under the marginal structural Cox model \eqref{eqn:mscm} and Assumptions \ref{assumption:independence}--\ref{assumption:no-interaction}, we have
        \begin{equation}
\label{eqn:Lambda_0}
{\Lambda}_0(t) = \int_0^t \dfrac{E\{ \omega_0(Z,X) dN(y)\}}{E\{\omega_0(Z,X)e^{\psi D} I(Y\geq y) \} }.
\end{equation}
\end{proposition}
Identification formula \eqref{eqn:Lambda_0} directly leads to a weighted version of the Breslow estimator \citep{breslow1972contribution}.

Second, it can be extended to allow for ignorable censoring. 
\begin{proposition}
\label{cor:ignorable-censoring}
    	Under the marginal structural Cox model \eqref{eqn:mscm}, Assumptions \ref{assumption:independence}--\ref{assumption:U},\ref{assumption:no-interaction} and condition \eqref{eqn:ignorable-censoring}, we have
	\begin{equation*}
	\exp(\psi) = \dfrac{ E \int dN(y) (-D) \widetilde{\omega}(Z,X,D,y)  \left\{m(1) \widetilde{\gamma}_1(y) - \widetilde{\gamma}_2^m(y) \right\} }
	{E \int dN(y) (1-D) \widetilde{\omega}(Z,X,D,y) \left\{m(0) \widetilde{\gamma}_1(y) - \widetilde{\gamma}_2^m(y) \right\} },
	\end{equation*}
	where  $\widetilde{\omega}(Z,X,D,y)  = \omega_0(Z,X) \left(\dfrac{D}{P(C(1) \geq y\mid X)} + \dfrac{(1-D)}{P(C(0) \geq y\mid X)} \right) $,  \\  $\widetilde{\gamma}_1(y) = E\left\{I(Y\geq y)\widetilde{\omega}(Z,X,D,y)\right\}$ and  $\widetilde{\gamma}_2^m(y) = E\left\{m(D) I(Y\geq y)\widetilde{\omega}(Z,X,D,y)  \right\}.$ If we additionally assume that $D\ind C(d) \mid X,$ then $P(C(d)\geq y \mid X) = P(C \geq y\mid X, D=d).$ 
\end{proposition}

Third, it can be extended to identify parameters in the conditional structural Cox model:
\begin{equation}
\label{eqn:mscm_cond}
\lambda^T_d(t\mid X) = \lambda^T_0(t\mid X) e^{ \psi d}.
\end{equation}


\begin{proposition}
\label{prop:identification_torben}
   	Under the conditional structural Cox model \eqref{eqn:mscm_cond}, Assumptions \ref{assumption:independence}--\ref{assumption:U},\ref{assumption:no-interaction} and condition \eqref{eqn:ignorable-censoring}, 
	\begin{equation*}
	\exp(\psi) = \dfrac{ E \int dN(y) (-D) 
\overline{\omega}(Z,X) \left\{m(1) \overline{\gamma}_1(y,X) - \overline{\gamma}_2^m(y,X) \right\} }
	{E \int dN(y) (1-D) \overline{\omega}(Z,X) \left\{m(0) \overline{\gamma}_1(y,X) - \overline{\gamma}_2^m(y,X) \right\} },
	\end{equation*}
	where 	$\overline{\omega}(Z,X) = \overline{h}(X) (2Z-1)/\left\{f(Z\mid X)\right\},$ $\overline{\gamma}_1(y,X) = E\left\{I(Y\geq y)\overline{\omega}(Z,X) \mid X\right\}, \overline{\gamma}_2^m(y,X) = E\left\{m(D) I(Y\geq y)\overline{\omega}(Z,X) \mid X \right\},$ 
  $m(1)\neq m(0)$ and $\overline{h}(X)$ is any non-zero measurable function of $X$.
\end{proposition}

Fourth, it can be extended to accommodate the case of competing risks. Let $(T(d),\epsilon(d))$ denote the potential time to one of the competing events, where $\epsilon(d) \in \{1,\cdots,K\}$ keeps track of which of the $K$ competing events would happen under exposure $d$. Let $S_d^{T,k}(t) = P(T(d)\geq t, \epsilon(d) = k)$ and $\lambda_d^{T,k}(t) = \lim\limits_{d t\rightarrow 0} {P(t \leq T(d) < t+dt, \epsilon(d)=k\mid T(d)\geq t)}/{dt}$ be the corresponding cause-specific survival and hazard functions, and $N^k(t) = I(Y\leq t, \Delta=1, \epsilon=k)$ be the cause-specific counting process. The cause-specific marginal structural Cox model  is:
\begin{equation}
    \label{eqn:cox-cause-specific}
    \lambda_d^{T,k}(t) = \lambda_0^{T,k}(t) e^{\psi_k d}, k=1,\ldots,K.
\end{equation}
\begin{proposition}
\label{prop:competing}
        Suppose that the cause-specific marginal structural Cox model \eqref{eqn:cox-cause-specific}, Assumptions \ref{assumption:independence}, \ref{assumption:relevance}, \ref{assumption:no-interaction} and the following conditions hold:
        \begin{itemize}
            \item[A3$^{*}$] $(T(d), \epsilon(d), C(d)) \ind (D,Z) \mid (X,U);$
            \item[A4$^{*}$] $C(d) \ind (T(d),\epsilon(d)), d=0,1.$
        \end{itemize}
        Then we have
	\begin{equation*}
	\exp(\psi_k) = \dfrac{ E \int dN^k(y) (-D) \omega_0(Z,X) \left\{m(1) \gamma_1(y) - \gamma_2^m(y) \right\} }
	{E \int dN^k(y) (1-D) \omega_0(Z,X) \left\{m(0) \gamma_1(y) - \gamma_2^m(y) \right\} },
	\end{equation*}
	where $m(D),\gamma_1(y), \gamma_2^m(y)$ satisfy the same conditions as in Theorem \ref{thm:identification_torben}.
\end{proposition}

Fifth, in the case of rare events, the no $U-Z$ interaction assumption \ref{assumption:no-interaction} may be replaced by a no $U-d$ interaction on the  hazard ratio scale: 
\begin{equation}
    \label{eqn:rare}
    \lambda_d^T(t\mid X,U) =  \lambda_0^T(t\mid X,U) e^{\beta(X) d}.
\end{equation}

\begin{proposition}
\label{prop:ind}
        Suppose that condition \eqref{eqn:rare}, 
        Assumptions \ref{assumption:independence}--\ref{assumption:U}, and the following conditions hold:
     \begin{itemize}
     \item[A4**] (Independent censoring) $T(d) \ind C(d) \mid X,U,$ and $Z \ind C \mid X, U$;
         \item[A5*] (Rare event)  $ S^T_d(y\mid X,U) = P(T(d)\geq y\mid X,U) \approx 1$ for all $y$ in a finite follow-up period.
     \end{itemize}   
       Then we have
    \begin{flalign*}
    \exp(\beta(X)) \approx \dfrac{\int E[ d N(y) (-D) \dfrac{2Z-1}{f(Z\mid X)} \mid X]  }{\int E[ d N(y) (1-D) \dfrac{2Z-1}{f(Z\mid X)} \mid X] }.
    \end{flalign*}
In particular,  if $\beta(X)$ is a constant function of $X$, then the conditional Cox model  \eqref{eqn:rare} approximates the marginal Cox model \eqref{eqn:mscm}  with $\psi = \beta(X).$ In this case, 
\begin{flalign*}
    \exp(\psi) \approx \dfrac{\int E[ d N(y) (-D) \dfrac{2Z-1}{f(Z\mid X)} h_1(X)]  }{\int E[ d N(y) (1-D) \dfrac{2Z-1}{f(Z\mid X)} h_1(X)] }
    \end{flalign*}
    for any non-zero measurable function $h_1(X).$
\end{proposition}

Proposition \ref{prop:ind} may also be extended to accommodate competing risks.  
\begin{corollary}
\label{cor:rare}
    If one assumes that 
    \begin{equation}
    \label{eqn:rare2}
    \lambda_d^{T,k}(t\mid X,U) =  \lambda_0^{T,k}(t\mid X,U) e^{\beta_k(X) d}, k=1,\ldots, K.
\end{equation}
Then under Assumptions \ref{assumption:independence}--\ref{assumption:U}, A4**, A5*,  and the assumption that $Z \ind Y(d), \epsilon(k) \mid X,U,$ we have
 \begin{flalign*}
    \exp(\beta_k(X)) \approx \dfrac{\int E[ d N^k(y) (-D) \dfrac{2Z-1}{f(Z\mid X)} \mid X]  }{\int E[ d N^k(y) (1-D) \dfrac{2Z-1}{f(Z\mid X)} \mid X] }.
\end{flalign*}
If $\beta_k(X),k=1,\ldots, K$ are constant  functions of $X$, then the conditional cause-specific Cox models \eqref{eqn:rare2} approximate the marginal cause-specific Cox model \eqref{eqn:cox-cause-specific} with $\psi_k = \beta_k(X)$ and 
\begin{flalign*}
    \exp(\psi_k) \approx \dfrac{\int E[ d N^k(y) (-D) \dfrac{2Z-1}{f(Z\mid X)} h_1^k(X)]  }{\int E[ d N^k(y) (1-D) \dfrac{2Z-1}{f(Z\mid X)} h_1^k(X)] }
    \end{flalign*}
    for any non-zero measurable functions $h_{1}^k(X),k=1,\ldots,K.$
\end{corollary}


\subsection{Estimation}
\label{subsec:Large sample properties}


In Theorem \ref{thm:identification_torben}, a natural choice for $m(D)$   is  $m(D) = D$. 
Under the modeling assumptions described in Section \ref{subsec:e.e}, \eqref{eqn:identification_torben} gives rise to the following estimator:
\begin{equation}
\label{eqn:torben}
\widehat{\psi} = \log \dfrac{ \sum\limits_{i=1}^n \Delta_i D_i \widehat{\omega}_0(Z_i,X_i)\left\{ { \widehat{\gamma}_{1,i} }-{\widehat{\gamma}_{2,i}^{m_0} }\right\} }
{ \sum\limits_{i=1}^n \Delta_i (1-D_i) \widehat{\omega}_0(Z_i,X_i) {\widehat{\gamma}_{2,i}^{m_0}} },
\end{equation}
where $\widehat{\omega}_0(Z_i,X_i) = (2Z_i-1)/\left\{f\left(Z_i|X_i; \widehat{\eta}\right) \delta^D \left( X_i; \widehat{\beta}\right) \right\}$, $\widehat{\gamma}_{1,i}=\widehat{\gamma}_{1}(Y_i)$, $\widehat{\gamma}_{2,i}^{m_0}=\widehat{\gamma}_{2}^{m_0}(Y_i)$ with
\begin{flalign*}
\widehat{\gamma}_{1}(y) =	n^{-1}\sum\limits_{j=1}^n I(Y_j \geq y)\widehat{\omega}_0(Z_j,X_j),
\widehat{\gamma}_{2}^{m_0}(y) = n^{-1}\sum\limits_{j=1}^n    D_j I(Y_j \geq y)%
\widehat{\omega}_0(Z_j,X_j). 
\end{flalign*}

We now discuss large sample properties for our proposed estimator \eqref{eqn:torben}.  Note that  $\widehat{\psi}$ solves the equation $\Pn \left\{H(\psi,\widehat\theta)\right\}=0$, where  $\theta=(\beta, \eta)$ and	
\begin{align*} 
H(\psi,\widehat\theta)=  \int\left [\{\widehat{\gamma}_{1}(y) - \widehat{\gamma}_{2}^{m_0}(y)\}D-(1-D)\widehat{\gamma}_{2}^{m_0}(y)\right ]\widehat{\omega}_0(Z,X)e^{-\psi D}dN(y)
\end{align*} 
It follows from \citet[][Lemma 5.10]{van2000asymptotic} that $\widehat\psi$ is a consistent estimator of $\psi_0$.
We may further write
$
\Pn H(\psi_0,\theta_0)= \Pn H^c(\psi_0,\theta_0)+ o_p(1/\sqrt{n}),
$
where 
$$
H^c(\psi_0,\theta_0)= \int\left[\left\{{\gamma}_{1}(y) - {\gamma}_{2}^{m_0}(y)\right\}D-(1-D){\gamma}_{2}^{m_0}(y)\right]\omega_0(Z,X)\left\{e^{-\psi D} dN(y) - R(y) d\Lambda_0(y)\right\}
$$
with
$R(y)=I(Y\geq y)$ and $\Lambda_0(y)=\int_0^y\lambda_0(s)\, ds$.
The $H^c_i(\psi_0,\theta_0)$'s are zero-mean  terms that are independent and identically distributed. We hence have the following theorem.

\begin{theorem}
    Suppose that the marginal structural Cox model \eqref{eqn:mscm} and the nuisance models $f(Z\mid X;\eta), \delta^D(X;\beta), OP(X; \zeta)$ are correctly specified. Under Assumptions \ref{assumption:independence} -- \ref{assumption:no-interaction}, we have that $\psi$ is asymptotically linear with influence function  given by
$
IF_{\widehat{\psi}} = 		-E\left\{ \left. \partial H(\psi, \theta_0)/\partial \psi \right|_{\psi = \psi_0}\right\}^{-1} \widetilde{H}(\psi_0, \theta_0),
$
where 
$
\widetilde{H}(\psi_0, \theta_0) = H^c(\psi_0,\theta_0) + E\left\{\left. \dfrac{\partial H(\psi_0, \theta)}{\partial \theta} \right|_{\theta = \theta_0} \right\} IF_{\widehat{\theta}}
$
with $IF_{\widehat{\theta}}$ being the influence function of $\widehat\theta$.
\end{theorem}


A consistent estimator of $n\mbox{var}(\widehat\psi)$ is
$
\mathbb{P}_n \{\widehat{IF}_{\widehat{\psi}}\}^2,
$
where $\widehat{IF}_{\widehat{\psi}}$ is obtained from $IF_{\widehat{\psi}}$ by replacing unknown quantities with their empirical counterparts.

\begin{remark}
    When the sample size is small, it is possible that \eqref{eqn:torben} is undefined as the term inside the logarithm is non-positive. 
\end{remark}

\section{Simulation studies}
\label{sec:simulation}

We now compare the finite sample performance of our proposed estimator $\widehat{\psi}$ to various other estimators proposed in the literature. In our simulations,   the baseline covariates $X$ include an intercept, a continuous variable $X_2$ generated from an exponential distribution with mean $1/\lambda_2$ and $X_3=X_2I(X_2\geq 1) - (X_2+1) I(X_2<1)$. These choices and the generating models below ensure that $\delta^D(X)$ is bounded away from 0, so that the instrumental relevance assumption holds.   The unmeasured confounder $U$ is generated from an independent exponential distribution with mean $1/\lambda_1$. Conditional on $X$ and $U$, the instrument $Z$ and treatment $D$ are generated from the following models:
$P(Z=1\mid X) = \text{expit}(- 1/\lambda_2 + X_2), 
\delta^D(X,U) = \text{tanh}(\beta_0 + \beta_1 X_2 + \beta_2 X_3 + \beta_3 U),$ and
$\log(OP^D(X,U)) = \zeta_0 + \zeta_1 U +  \zeta_2 X_2,$
where $\lambda_1 = \lambda_2 = 2, \zeta_0=-2, \zeta_1=\zeta_2=1$. We let $(\beta_0,\beta_1,\beta_2) = (0.5, 0.5,0) \text{ or } (0,0,0.5)$. Moreover, the first set of parameter values is compatible with the commonly used monotonicity assumption that $D(1)\geq D(0)$ almost surely, as $\delta^D(X)$ is always positive. The censoring time $C$ was generated from an exponential distribution with mean $1/\lambda$.   As discussed in detail in \cite{richardson2017modeling}, our specifications of $\delta^D(X)$ and $\log(OP^D(X))$ give rise to  a unique model on $P(D=1\mid Z=z, X), z=0,1.$ Visualizations of such a model can be found in \citet[][Supplementary Materials, upper panels of Figure 1]{richardson2017modeling}.
To make the observed data models compatible with a marginal structural Cox model with parameter $\psi$, as explained in the Supplementary Material, we let the survival outcome $T$ be the unique root of the following function \citep{tchetgen2012parametrization}:
$$
f(t) = 	\dfrac{1}{\lambda_1} (\lambda_1 - \kappa_1 t) \dfrac{1}{\lambda_2} (\lambda_2 - \kappa_2 t)  \exp\left\{(\kappa_1 U + \kappa_2 X_2 - \lambda_0 e^{\psi D}) t \right\} - 1 + A,
$$
where $\psi=0.5, \lambda_0  =4, \kappa_1 = \kappa_2 = 1$ and $A$ is uniformly distributed on the interval $[0,1].$ 

In addition to the proposed estimator \eqref{eqn:torben}, we also implement the following estimators: (i) {\sf Cox-crude}: a crude Cox proportional model not adjusting for any covariates; (ii)  {\sf Cox-adj}: a Cox model adjusting for covariates $(X_2, X_3)$; (iii) {\sf Cox-MSM}: marginal structural Cox model  adjusting for  $(X_2,X_3)$; (iv) {\sf MacKenzie}: \cite{mackenzie2014using}'s method; (v) {\sf TSPS}: a naive application of the two stage least square method, with a first stage linear model and a second stage Cox model; (vi) {\sf TSRI}:  a naive application of the two stage residual inclusion method \citep{terza2008two}, with a first stage linear model and a second stage Cox model.


\begin{table}[!htbp] 
	\begin{center}
	\small
		\caption{{{Censoring rate and bias times 100 (standard error times 100, in parenthesis)  for various methods estimating  the log causal hazard ratio $\psi$. The true value for $\psi$ is 0.5.  Here ``monotonicity holds'' refers to the case where $\delta^D(X) > 0$ for all $X$. The sample size is 1000. Results are based on 1000 simulated data sets}}}
		\label{tab:results}
		\begin{tabular}{rccccccccccc}
			& \multicolumn{1}{c}{Censoring \%} &  \multicolumn{6}{c}	{Bias$\times 100$  (SE $\times 100$)} \\[5pt]
				\multicolumn{2}{l}{A\ref{assumption:no-interaction} holds ($\beta_3$=0)}	&  Proposed & {\sf Cox-crude} & {\sf Cox-adj} &  {\sf Cox-MSM} & {\sf MacKenzie} & {\sf TSPS} & {\sf TSRI} \\[5pt]
			\multicolumn{4}{l}{Monotonicity holds} && \\[2pt]
		$\lambda=0$ &      0 &  0.41(0.44) & 5.4(0.22) & 1.5(0.23) &  2.5(0.27) &  5.7(0.35) & $-$3.9(0.35) & 0.26(0.35) \\
$\lambda=1$ & 17.2\% &  0.47(0.48) & 5.5(0.24) & 1.5(0.25) &  2.6(0.29) &  5.9(0.38) & $-$2.4(0.39) & 0.20(0.39) \\
$\lambda=4$ & 44.9\% &  0.97(0.58) & 5.7(0.29) & 1.8(0.30) &  2.6(0.34) &  6.7(0.44) & 0.19(0.47) & 0.65(0.47) \\[8pt]
		\multicolumn{4}{l}{Monotonicity fails} && \\[2pt]
	$\lambda=0$ &         0 & $-$0.21(0.45) & 3.1(0.21) & 1.7(0.22) & $-$3.8(0.30) & $-$7.4(0.53) & $-$8.9(0.47)& -3.2(0.49) \\
$\lambda=1$ & 17.0\% & $-$0.38(0.49) & 3.2(0.23) & 1.7(0.23) & $-$4.0(0.32) & $-$7.5(0.58) & $-$7.4(0.53)& -2.9(0.54) \\
$\lambda=4$ & 44.6\% & $-$0.65(0.60) & 2.9(0.28) & 1.2(0.28) & $-$4.3(0.39) & $-$8.8(0.69) & $-$6.2(0.64)& -3.6(0.65)  \\[15pt]
		\multicolumn{3}{l}{A\ref{assumption:no-interaction} fails ($\beta_3$ = 0.5)}  &  &  &   & &  \\[5pt]
			\multicolumn{4}{l}{Monotonicity holds} && \\[2pt]
	$\lambda=0$ &        0 & $-$0.31(0.36) & 5.1(0.22) & 1.2(0.22) &  2.1(0.30) & 4.6(0.30) &  $-$4.0(0.30) & 0.00(0.29) \\
$\lambda=1$ & 17.2\% & $-$0.29(0.39) & 5.2(0.24) & 1.3(0.24) &  2.2(0.32) & 4.7(0.33) &  $-$2.7(0.33) & 0.02(0.33) \\
$\lambda=4$ & 44.8\% &  0.06(0.47) & 5.6(0.28) & 1.6(0.29) &  2.4(0.38) & 5.6(0.38) & $-$0.22(0.40)& 0.50(0.40) \\[8pt]
		\multicolumn{4}{l}{Monotonicity fails} && \\[2pt]
$\lambda=0$ &        0 &   2.5(0.72) & 3.7(0.20) & 2.1(0.21) & $-$3.3(0.30) & $-$10(0.97) &  $-$8.1(0.87)  & -3.7(0.89)\\
$\lambda=1$ & 17.2\% &   2.5(0.80) & 3.9(0.22) & 2.2(0.22) & $-$3.4(0.32) &  $-$11(1.1) &  $-$7.0(0.97) & -3.7(0.99) \\
$\lambda=4$ & 44.8\% &   2.6(0.97) & 3.8(0.27) & 2.0(0.27) & $-$3.4(0.40) &  $-$14(1.3) &   $-$7.2(1.2) &  -5.1(1.2)\\[5pt]
		\end{tabular}
		\label{result2}
	\end{center}
\end{table}

\begin{table}[!htbp] 
	\begin{center}
		\caption{{{Range of censoring rate, bias times 100 (standard error times 100, in parenthesis) and coverage rate for the proposed method estimating  the log causal hazard ratio $\psi$.  The nominal coverage rate is 95\%. Here ``monotonicity holds'' refers to the case where $\delta^D(X) > 0$ for all $X$. The sample size is 1000. Results are based on 1000 simulated data sets}}}
		\label{tab:results4}
		\begin{tabular}{rcccrrrrrrr}
			&& \multicolumn{1}{c}{Censoring rate} &&  \multicolumn{2}{c}	{Bias$\times 100$  (SE $\times 100$)} && \multicolumn{2}{c}	{Coverage rate} \\[5pt]
				\multicolumn{2}{l}{Assumption \ref{assumption:no-interaction} holds} &  && $\psi = 0$ &  $\psi=0.5$ && $\psi = 0$ &  $\psi=0.5$ \\[5pt]
			\multicolumn{1}{l}{Monotonicity holds} &&&&& \\[2pt]
			$\lambda=0$ &  &                0 &  & $-$0.43(0.41) & $-$0.26(0.43) &  & 0.952 & 0.949 \\
			$\lambda=1$ &  & 17.2\%--20.0\% &  & $-$0.53(0.46) & $-$0.24(0.47) &  & 0.954 & 0.951 \\
			$\lambda=4$ &  & 44.8\%--50.0\% &  & $-$0.28(0.58) & $-$0.01(0.57) &  & 0.961 & 0.956 \\[8pt]
			\multicolumn{1}{l}{Monotonicity fails} &&&&& \\[2pt]
			$\lambda=0$ &  &                0 &  &  0.37(0.43) &  0.41(0.44) &  & 0.945 & 0.951 \\
			$\lambda=1$ &  & 17.0\%--20.0\% &  &  0.52(0.48) &  0.56(0.48) &  & 0.940 & 0.953 \\
			$\lambda=4$ &  & 44.5\%--50.0\% &  &  0.61(0.60) &  0.73(0.59) &  & 0.950 & 0.956 \\[20pt]
			\multicolumn{2}{l}{Assumption \ref{assumption:no-interaction} fails} 	 &  && $\psi = 0$ &  $\psi=0.5$ && $\psi = 0$ &  $\psi=0.5$ \\[5pt]
			\multicolumn{1}{l}{Monotonicity holds} &&&&& \\[2pt]
			$\lambda=0$ &  &                0 &  & $-$0.19(0.34) & $-$0.69(0.35) &  & 0.952 & 0.946 \\
			$\lambda=1$ &  & 17.1\%--20.0\% &  & $-$0.27(0.37) & $-$0.69(0.38) &  & 0.955 & 0.948 \\
			$\lambda=4$ &  & 44.7\%--50.0\% &  & $-$0.14(0.47) & $-$0.61(0.46) &  & 0.956 & 0.955 \\[8pt]
			\multicolumn{1}{l}{Monotonicity fails} &&&&& \\[2pt]
			$\lambda=0$ &  &                0 &  &  0.34(0.67) &   3.2(0.70) &  & 0.951 & 0.960 \\
			$\lambda=1$ &  & 17.1\%--20.0\% &  &  0.71(0.76) &   3.8(0.79) &  & 0.948 & 0.960 \\
			$\lambda=4$ &  & 44.7\%--50.0\% &  &    1.2(1.0) &    4.8(1.0) &  & 0.963 & 0.966 \\[5pt]
		\end{tabular}
	\end{center}
\end{table}

All simulation results are based on 1000 Monte-Carlo runs of n = 1000 units each. Table \ref{tab:results} summarizes the simulation results. When $\beta_3 = 0$ such that Assumption \ref{assumption:no-interaction} holds, 
the biases from Cox regression estimates {\sf Cox-crude}, {\sf Cox-adj}  and {\sf Cox-MSM} are large, due to unmeasured confounding by $U$. 
\cite{mackenzie2014using}'s method, {\sf TSPS} and {\sf TSRI}  are also biased, while
the bias of the proposed estimator is small relative to its standard deviation; see also Table \ref{tab:results5} in the Supplementary Material. Consistent with previous results in the literature \citep{wan2018general}, the bias of {\sf TSRI} is in general smaller than that of the {\sf TSPS}. Results in Table \ref{tab:results4}  shows that Wald-type 95\% confidence intervals constructed using the proposed variance estimator also achieve the nominal coverage rate in all the scenarios under which Assumption \ref{assumption:no-interaction} holds, confirming our theoretical results.
Given a fixed data generating mechanism for the censoring time $C$, the censoring rate only increases with  $\psi$ slightly. When $\beta_3 = 0.5$ so that Assumption \ref{assumption:no-interaction} fails to hold, \cite{mackenzie2014using}'s estimator produces an invalid estimate in one of the 1000 Monte Carlo runs; all the other estimators produce valid estimates in all Monte Carlo runs.  The proposed estimator $\widehat{\psi}$ has a large bias only when the monotonicity condition fails and $\psi=0.5$. The 95\% confidence intervals, however, are only slightly conservative. For example, when $\psi=0.5, \lambda=4$ and monotonicity fails, although the bias of $\widehat{\psi}$ (4.8) is much larger than its standard error (1.0), it is much smaller compared to its standard deviation (31.6). As a rule of thumb, the performance of interval estimates begins to deteriorate when the bias is more than 40\% of standard deviation \citep[e.g.][]{kang2007demystifying}. So it is not surprising to see that in this case, the coverage probability, 96.6\%, is only slightly larger than the nominal level.



\section{Application to the Health Insurance Plan Study}

In this section, we illustrate the proposed method by revisiting the Health Insurance Plan study, a randomized trial of mammography screening from 1963 to 1982. The goal was to determine whether screening reduced breast cancer mortality in women. 60,695 women aged between 40 and 60 were randomized into two groups. Half of them, in the study group, were assigned to receive two annual breast examinations that include mammography, a breast exam and an interview. The control group continued to receive their usual care. About 35\% of women offered screening (9984 out of 30130) refused to participate, so there was a significant portion of non-compliers. Furthermore, study women with a higher risk for breast cancer tended to comply: the incidence rate among study group women who refused screening was 1.45 per 1,000, versus 1.87 per 1000 among control group women. So a direct comparison between the women who accepted screening, and women who did not receive screening, is subject to unmeasured confounding. The same data were used by \cite{joffe2001administrative} to estimate the causal effect under an accelerated failure time model, and \cite{martinussen2017instrumentalcumulative} to estimate the conditional causal hazard difference. Instead, we shall use the proposed method to estimate the marginal causal hazard ratio due to mammography screening. 

 Following previous analyses by \cite{joffe2001administrative} and \cite{martinussen2017instrumentalcumulative}, we focus on the first 10 years of follow-up to reduce attenuation of the effects of the screening. We consider the randomization variable as our instrument $Z$, and the indicator of receiving screening as our exposure $D$. Our primary outcome of interest is breast cancer mortality. For verification purposes, we also consider a secondary outcome, death due to other causes, for which we expect the causal effect of breast cancer screening to be null. In the first 10 years of follow-up,
there were 4221 deaths but only 340 were deemed due to breast cancer.  Note that the independence and instrumental relevance assumptions hold by design as the instrument is randomized and only subjects assigned to the treatment group may receive screening.  The exclusion restriction   is  plausible because randomization to the study group is unlikely to affect mortality directly, had a study women chosen to refuse screening.  We adjust for baseline covariate age, as a predictor of compliance behavior. 
 
We then use the proposed methods in Section \ref{sec:identification} to estimate the marginal causal hazard ratio. To accommodate the two competing outcomes we consider here, we shall apply the results in Proposition \ref{prop:competing} (denoted as {\sf Proposed}). In addition, since the outcomes are relatively rare, we also apply Corollary \ref{cor:rare} (denoted as {\sf Proposed-rare-event}). In doing so, we assume that cause-specific hazard ratios $\beta_k(X),k=1,2$ are constant functions of $X$, so that it targets the same parameters as {\sf Proposed}. We also assume that $h_1^k(X) = 1, k=1,2$ in Corollary \ref{cor:rare}. For comparison purpose, we also implement the estimators {\sf Cox-crude}, {\sf Cox-adj}, {\sf Cox-MSM}, {\sf MacKenzie}, {\sf TSPS} and {\sf TSRI}, in which except for  {\sf Cox-crude} and {\sf MacKenzie}, we adjust for the baseline covariate age.


\begin{table}[!htbp] 
	\begin{center}
		\caption{Point estimates (95\% CI) for hazard ratio of breast cancer screening on death due to different reasons} 
		\bigskip
		\begin{tabular}{rcc}
			\toprule
		Method &  Death due to breast cancer &  Death due to other causes \\
			\midrule
			{\sf {Cox-crude}} & 0.77 (0.61,0.97) & 1.40 (1.30, 1.50) \\[2pt]
			{\sf {Cox-adj}} & 0.77 (0.61,0.97) &  1.37 (1.28, 1.47)             \\ [2pt]
			{\sf {Cox-MSM}} & 0.79 (0.62, 1.01) & 0.66 (0.61, 0.71) \\[2pt]
			{\sf {MacKenzie}} & 0.66 (0.46, 0.94)  & 0.99 (0.90, 1.08) \\[2pt]
			{\sf {TSPS}} & 0.66 (0.48, 0.92) & 0.99 (0.90, 1.09) \\[2pt]
			{\sf {TSRI }} & 0.67 (0.45, 0.99)  & 0.96 (0.86,1.07) \\[2pt]
			{\sf Proposed} & 0.67 (0.50,0.89) & 0.99 (0.90, 1.09) \\[2pt] 
			{\sf Proposed-rare-event} & 0.68 (0.51,0.90) & 1.02 (0.91, 1.15) \\
			\bottomrule
		\end{tabular}
		\label{tab:result2}
	\end{center}
\end{table}

Table \ref{tab:result2} summarizes the results. 
The crude and adjusted Cox regression model and the marginal structural Cox model all suggest that breast cancer screening is negatively associated with breast cancer mortality (hazard ratio: 0.77, 0.77, 0.79). Such associations, however, may be distorted by the fact that the screening group is at higher risk for breast cancer compared to the group that did not receive screening. Based on this reason, one would expect that the effect of breast cancer screening would be stronger than what these associations suggested.  Indeed, our analysis based on the identification formula outlined in Proposition \ref{prop:competing} suggests that breast cancer screening reduces the hazard of death due to breast cancer, with a hazard ratio of 0.67 (95\% CI = [0.50,0.89]). Furthermore,   as expected, breast cancer screening has little effect for deaths of reasons other than breast cancer, with a hazard ratio of 0.99 (95\% CI = [0.90, 1.09]).  Our proposed methods assuming rare events and no $D-U$ interaction in the Causal Cox model,  yield very similar results to that assuming no $Z-U$ interaction in an additive model for the treatment.

Both {\sf MacKenzie} and the two-stage methods {\sf TSLS} and {\sf TSRI} yield similar point estimates to our proposed estimator, with slightly larger variances.   In this application, the random assignment to the study group is a valid instrumental variable even without conditioning on age, so the unconditional IV model assumed by  {\sf MacKenzie} is also valid. Moreover,  the outcomes considered in this example are rare, in which case both  {\sf MacKenzie} and the two-stage methods are known to be approximately unbiased \citep{mackenzie2014using,tchetgen2015instrumental}. These results confirm the findings from our primary analysis based on the identification formula outlined in Proposition \ref{prop:competing}.

\section{Discussion}
\label{sec:discussion}

In this article, we considered the identification and estimation of the marginal causal hazard ratio under the proportional hazards assumption.
Our framework can also be extended in the following directions. First, in longitudinal studies, it is often the case that both the treatment and confounding variables are time dependent. It would be interesting to extend the proposed methods to estimate parameters in a marginal structural Cox model with time-varying treatments. {This has been done for the effect of treatment among the treated under the Cox regression model}  \citep{martinez2019instrumental} {but not for the marginal causal hazard ratio which is defined under a marginal structural Cox model}. Second, with an uncensored outcome, one can construct a locally efficient estimator for the population treatment effect of interest that is also multiply robust in the sense that such an estimator is consistent in the union of three different observed data models \citep{wang2018bounded}. Deriving a locally semiparametric efficient estimator for the causal hazard ratio under our identification assumptions is an important venue for future research.  Third, so far we have restricted our analysis to a single binary instrument and a binary exposure. It has been shown that the framework in \cite{wang2018bounded} can be extended to allow for general instruments and exposure \citep{hartwig2020average}. We leave this as   future work to extend the proposed method to the case of general instruments and exposure.


\section*{Acknowledgements}
Wang was supported by the Natural Sciences and Engineering Research Council of Canada. Tchetgen Tchetgen was supported by  the National Institutes of Health. Wang is also affiliated with the Department of Computer and Mathematical Sciences,  University of Toronto Scarborough.
Vansteelandt is also affiliated with the Department of Medical Statistics at the London School of Hygiene and Tropical Medicine, UK.

\section*{Data Availability Statement}

The data that support the findings of this paper are available on request from the corresponding author. The data are not publicly available due to privacy or ethical restrictions.




	\thispagestyle{empty}
\bibliographystyle{apalike}
\bibliography{causal}

\section*{Supporting Information}

Tables,
Figures, Proofs of theorems, propositions and claims referenced in
Sections \ref{sec:framework}, \ref{sec:identification} and \ref{sec:simulation}   are available with this paper. 
The {\tt R} programs that were used to analyze the data can be obtained from \url{https://doi.org/10.7910/DVN/FL4KFL}.

\clearpage

 \centerline{\large\bf Supplementary Material for ``Instrumental variable estimation of}
\vspace{2pt}
 \centerline{\large\bf causal hazard ratio''}
\vspace{.25cm}
\vspace{.4cm} \centerline{Linbo Wang, Eric Tchetgen Tchetgen, Torben Martinussen, Stijn Vansteelandt} \vspace{.4cm}  \vspace{.55cm} 
\par

\setcounter{equation}{0}
\setcounter{figure}{0}
\setcounter{table}{0}
\setcounter{section}{0}
\renewcommand{\theequation}{S\arabic{equation}}
\renewcommand{\thefigure}{S\arabic{figure}}
\renewcommand{\thetable}{S\arabic{table}}
\renewcommand{\theassumption}{S\arabic{assumption}}
\def\thesection{S\arabic{section}}

	\begin{abstract}
	The supplementary file contains additional causal diagrams, simulation results, and proofs for all the theorems and propositions.
	\end{abstract}

\section{Alternative causal diagram}

Figure \ref{DAG:iv_model2} provides an alternative causal diagram that is compatible with the  instrumental variable assumptions \ref{assumption:independence} -- \ref{assumption:ind_censoring}.

\begin{figure}[!htbp]
	\parbox{.4\textwidth}{
		\vspace{0.9cm}
		\centering
		\begin{tikzpicture}[->,>=stealth',node distance=1cm, auto,]
		\node[est] (Z) {$Z$};
		\node[est, right = of Z] (D) {$D$};
		\node[est, right = of D] (T) {$T$};
		\node[est, below = of D] (X1) {$X_1$};
		\node[est, above = of D] (V) {$X_2$};
		\node[est, above = of T] (C) {$C$};
		\node[shade, below = of T] (U) {$U$};
		\path[pil] (Z) edgenode {} (D);
		\path[pil] (D) edgenode {} (T);
 		\path[pil] (U) edgenode {} (D);
 		\path[pil] (U) edgenode {} (T);
		\path[pil] (V) edgenode {} (Z);
		\path[pil] (V) edgenode {} (D);
		\path[pil] (V) edgenode {} (T);
			\path[pil] (Z) edgenode {} (X1);
		\path[pil] (X1) edgenode {} (D);
		\path[pil] (X1) edgenode {} (T);
		\path[pil] (D) edgenode{} (C);
		\end{tikzpicture}
		\quad \\ \bigskip (a). A directed  acyclic graph.
	}
	\parbox{.4\textwidth}{
		\centering	
		\begin{tikzpicture}[>=stealth, ultra thick, node distance=2cm,
		pre/.style={->,>=stealth,ultra thick,black,line width = 1.5pt}]
		\begin{scope}
		\node[name=Z,shape=ellipse splitb, ellipse splitb/colorleft=black, ellipse splitb/colorright=red, 
		ellipse splitb/innerlinewidthright = 0pt,  
		/tikz/ellipse splitb/linewidthright = 1pt,   
		ellipse splitb/gap=3pt, style={draw},rotate=90] {
			{\rotatebox{-90}{$Z$\;}}
			\nodepart{lower}
			{\rotatebox{-90}{\;\textcolor{red}{$z$}}}
		};
		\node[name=D,shape=ellipse splitb, below right=0.65cm and 1.8cm of Z,  ellipse splitb/colorleft=black, ellipse splitb/colorright=red, 
		ellipse splitb/innerlinewidthright = 0pt,  
		/tikz/ellipse splitb/linewidthright = 1pt,   
		ellipse splitb/gap=3pt, style={draw},rotate=90] {
			{\rotatebox{-90}{$D(\tred{z})$\;}}
			\nodepart{lower}
			{\rotatebox{-90}{\;\textcolor{red}{$d$}}}
		};
		\node[thick, name=T,shape=ellipse,style={draw}, below right =0.05cm and 1.8cm  of D,outer sep=0pt, text width = 8mm]
		{$T({\color{red}{d}})$
		};
		\node[thick, name=C,shape=ellipse,style={draw}, above  =1.2cm  of T, outer sep=0pt, text width = 8mm]
		{$C({\color{red}{d}})$
		};
	\node[shade, below =1.2cm of T] (U) {$U$};
		\draw[pil,->] (Z) to (D);
		\node[est, above right=1.5cm and 0.3cm of D] (V) {$X_2$};
		\node[est, below =3.5cm of V] (X1) {$X_1$};
		\path[pil] (Z) edgenode {} (D);
		\path[pil] (D) edgenode {} (T);
		\path[pil] (D) edgenode {} (C);
		\path[pil] (U) edgenode {} (T);
		\draw[pil, ->] (2.7,2)  to[bend right]  (-0.5,0.4);
		\draw[pil, ->] (3.1,1.7)  --  (3.1,0.4);
		\path[pil] (V) edgenode {} (T);
		\draw[pil, <-] (2.7,-2.2)  to[bend left]  (-0.5,-0.4);
		\draw[pil, ->] (3.1,-1.7)  --  (3.1,-0.4);
		\draw[pil, ->] (7.5,-1.7)  --  (3.2,-0.5);
			\draw[pil, ->] (3.7,-2.2) to[bend right]  (7.5,-0.5);
		\end{scope}
		\end{tikzpicture}
		\quad \\ \bigskip	(b). A single world intervention graph.
	}
	\caption{An alternative causal diagram  compatible with the  instrumental variable assumptions \ref{assumption:independence} -- \ref{assumption:ind_censoring}.  Variables $X,Z,D$ are observed, where $X=(X_1,X_2)$; $T$ is possibly right censored.  For simplicity, we do not include arrows between $X_1$ and $X_2$, but they can be arbitrarily related. The left panel gives a causal directed acyclic graph \citep{pearl2009causality}, and the right panel gives a single world intervention graph \citep{richardson2013single}.}
	\label{DAG:iv_model2}
\end{figure}

\section{Proof of Theorem \ref{thm:identification}}

Let $N_d(y) = I(Y(d) \leq y, \Delta(d) = 1), d=0,1.$
Note that $$ E\left\{ dN(y)\dfrac{2Z-1}{f\left( Z|X\right) \delta^D
	\left( X\right) }D\right\}  =  E_X\left[\dfrac{1}{\delta^D \left( X\right) } \sum\limits_{z=0,1}  (2z-1) E\left\{  dN(y) D|Z=z,X\right\} 
\right], $$ 
where 
\begin{flalign*}
&	\sum\limits_{z=0,1}  (2z-1) E\left\{  dN(y) D|Z=z,X\right\}  \\
&= 	 \sum\limits_{z=0,1}  (2z-1)  E\left\{  dN_1(y) D|Z=z,X\right\} \quad \text{(consistency)} \\
&=	 \sum\limits_{z=0,1}  (2z-1) E_{U\mid X} E\left\{ dN_1(y)		D|X,U,Z=z\right\}  \quad (Z\ind U\mid X) \\
&= 	 \sum\limits_{z=0,1}  (2z-1)  E_{U\mid X}\left[ E\left\{ dN_1(y)|X,U,Z=z\right\} E\left\{ D |X,U,Z=z\right\} \right]   \quad (D\ind T(1), C(1)\mid Z, X, U) \\
&= \sum\limits_{z=0,1}  (2z-1)  E_{U\mid X}\left[ E\left\{ dN_1(y)|X,U\right\} E\left\{D|X,U,Z=z\right\} \right]  \quad(Z\ind T(1),C(1)\mid U,X) \\
&= \delta^D(X) E_{U\mid X}\left[ E\left\{ dN_1(y)|X,U\right\}  \right]  \quad (\text{due to   \eqref{eqn:iden}})\\
&= \delta^D(X) E\left\{ dN_1(y)|X\right\}
\end{flalign*}
so that
\begin{flalign*}
E\left\{ dN(y)\dfrac{2Z-1}{f\left( Z|X\right) \delta^D
	\left( X\right) }D\right\}   &= E\left\{ dN_1(y)\right\} 
 \\ &= P(y\leq T(1) < y+dy, C(1)\geq y+dy)\\
 &= (dP(T(1) \leq y)) P(C(1) \geq y) \\
  &= \dfrac{dP(T(1) \leq y)}{S_1^T(y) dt} S_1^T(y) S_1^C(y) \\ &
= \lambda_1^T(y) S_1^Y(y) dy.
\end{flalign*}
Note similar to the proof above, we obtain that for any measurable function $H$,
\begin{flalign}
E\left\{ H(Y(1))\dfrac{2Z-1}{f\left( Z|X\right) \delta^D
	\left( X\right) }D\right\}   &= E\left\{ H(Y(1))\right\};  \label{result1}\\
E\left\{ H(Y(0))\dfrac{2Z-1}{f\left( Z|X\right) \delta^D
	\left( X\right) }(1-D)\right\}   &= -E\left\{ H(Y(0))\right\} \label{result0};\\
E\left\{ H(Y)\dfrac{2Z-1}{f\left( Z|X\right) \delta^D
	\left( X\right) }\right\}   &= E\left\{H(Y(1)) - H(Y(0))\right\} \label{eqn:result2}.
\end{flalign}
We shall use these equations repeatedly in the following proof.

Due to \eqref{eqn:result2},
\begin{equation}
\label{eqn:step}
E\left\{ dN(y)\dfrac{2Z-1}{f\left( Z|X\right) \delta^D
	\left( X\right) }\right\}  = E\left\{ dN_1(y)-dN_{0}(y)\right\} = \left\{\lambda_1^T(y) S_1^Y(y) - \lambda_0^T(y) S_0^Y(y)\right\}dy.
\end{equation}
Due to \eqref{result1},
\begin{flalign*}
&E\left\{ De^{\psi D} I(Y\geq y)\dfrac{2Z-1}{%
	f\left( Z|X\right) \delta^D \left( X\right) }\right\}  \\
&=e^{\psi} E\left\{ D I(Y(1)\geq y)\dfrac{2Z-1}{%
	f\left( Z|X\right) \delta^D \left( X\right) }\right\}  \\
&= e^\psi E\left\{I(Y(1)\geq y)\right\} 
= e^{\psi} S_1^Y(y).
\end{flalign*}%
Finally, due to \eqref{result1} and \eqref{result0},
\begin{flalign*}
&E\left\{ e^{\psi D} I(Y\geq y)\dfrac{2Z-1}{%
	f\left( Z|X\right) \delta^D \left( X\right) }\right\}  \\
&= e^\psi E\left\{D I(Y(1)\geq y)\dfrac{2Z-1}{%
	f\left( Z|X\right) \delta^D \left( X\right) }\right\}  + E\left\{(1-D) I(Y(0)\geq y)\dfrac{2Z-1}{%
	f\left( Z|X\right) \delta^D \left( X\right) }\right\} \\ 
&= e^\psi E\left\{I(Y(1))\geq y\right\} - E\left\{I(Y(0)\geq y)\right\} = e^\psi S_1^Y(y) - S_0^Y(y). 
\numberthis\label{eqn:step2}
\end{flalign*}%

Therefore 
\begin{small}
	\begin{flalign*}
	E\left\{ H\left( \psi \right) \right\}  
	&=\int \left[ E\left\{ dN(y)h(D)\dfrac{2Z-1}{f\left(
		Z|X\right) \delta^D \left( X\right) }D\right\} -\dfrac{ E\left\{ dN(y)h(D)\dfrac{2Z-1}{%
			f\left( Z|X\right) \delta^D \left( X\right) }\right\}E\left\{ De^{\psi D}h(D) I(Y\geq y)\dfrac{2Z-1}{f\left( Z|X\right) \delta^D
			\left( X\right) }\right\} }{E\left\{ e^{\psi D}h(D) I(Y\geq y)\dfrac{2Z-1}{f\left( Z|X\right) \delta^D
			\left( X\right) }\right\} }\right]  \\
	&=\int dy\left[ \lambda_1^T(y) S_1^Y(y)h(1) -   \dfrac{\left\{h(1)\lambda_1^T(y)S_1^Y(y) - h(0) \lambda_0^T(y) S_0^Y(y)\right\} h(1)e^\psi S_1^Y(y) }{h(1)e^\psi  S_1^Y(y) - h(0)S_0^Y(y)}\right] \numberthis\label{eqn:key} \\
	&=0,
	\end{flalign*}%
\end{small}
where $h(1)e^\psi  S_1^Y(y) - h(0)S_0^Y(y)\neq 0 a.e.$ as long as  $h(1)h(0)<0$. Furthermore, we have
\begin{flalign*}
\dfrac{\partial E\{H(\psi)\}}{\partial \psi} &=-\int  e^\psi  b \dfrac{\left\{h(1)\lambda_1^T(y)S_1^Y(y) - h(0) \lambda_0^T(y) S_0^Y(y)\right\} h(1) S_1^Y(y)h(0)S_0^Y(y) }{\left\{ h(1)e^\psi  S_1^Y(y) - h(0)S_0^Y(y)  \right\}^2} dy,
\end{flalign*}
the sign of which does not depend on $\psi$ and is non-zero as long as $h(1)h(0)<0$. Hence the solution to equation $E\{H(\psi)\}=0$ is unique.




		\section{Proof of Theorem \ref{thm:identification_torben}}

	We first note that 
	\begin{flalign*}
	    \gamma_1(y) &=  E\left\{I(Y\geq y)\omega_0(Z,X)\right\}     =  S_1^Y(y) - S_0^Y(y); \\
	    \gamma_2^m(y) &= E\left\{m(D) I(Y\geq y)\omega_0(Z,X)  \right\}
	    =  m(1) S_1^Y(y) - m(0)S_0^Y(y).
	\end{flalign*}
	Hence
	\begin{flalign*}
m(1) \gamma_1(y) - \gamma_2^m(y) 
	&=  S_0^Y(y) \left\{m(0)-m(1)\right\}; \\
m(0) \gamma_1(y) - \gamma_2^m(y) 
	&=  S_1^Y(y) \left\{m(0)-m(1)\right\}; \\
E  dN(y) (-D) \omega_0(Z,X) &= -  \lambda_1^T(y) S_1^Y(y); \\
E  dN(y) (1-D) \omega_0(Z,X) &= -  \lambda_0^T(y) S_0^Y(y).
	\end{flalign*}
Hence 
\begin{flalign*}
RHS = \dfrac{\int -  \lambda_1^T(y) S_1^Y(y)   S_0^Y(y) \left\{m(0)-m(1)\right\}  dy}{ \int  -  \lambda_0^T(y) S_0^Y(y)  S_1^Y(y) \left\{m(0)-m(1)\right\} dy } = exp(\psi).
\end{flalign*}

		\section{Proof of Proposition \ref{proposition:breslow}}

Due to \eqref{eqn:step} and \eqref{eqn:step2},
\begin{flalign*}
& \int_0^t \dfrac{E\{ \omega_0(Z,X) dN(y)\}}{E\{\omega_0(Z,X)e^{\psi D} I(Y\geq y) \} }  \\
&= \int_0^t \dfrac{\left\{\lambda_1^T(y) S_1^Y(y) - \lambda_0^T(y) S_0^Y(y)\right\}dy}{e^\psi S_1^Y(y) - S_0^Y(y)} \\
&= \int_0^t \lambda_0^T (y) dy = \Lambda_0^T(t).
\end{flalign*}	
		
			\section{Proof of Proposition \ref{cor:ignorable-censoring} }
	
	Let $\omega_c(D,X) = \dfrac{D}{P(C(1)\geq y\mid X)} + \dfrac{1-D}{P(C(0)\geq y\mid X)}.$
	Under \eqref{eqn:ignorable-censoring},  similar to the proof of Theorem \ref{thm:identification}, we have
	    \begin{flalign*}
E\left\{ dN(y)\dfrac{2Z-1}{f\left( Z|X\right) \delta^D
	\left( X\right) } \omega_c(D,X) D\right\}   &= E_X \dfrac{E\left\{ dN_1(y) \mid X\right\}}{P(C(1)\geq y\mid X)}
 \\ &= E_X \dfrac{ P(y\leq T(1) < y+dy, C(1)\geq y+dy \mid X) } {P(C(1)\geq y\mid X)} \\
 &= dP(T(1) \leq y) = \dfrac{dP(T(1) \leq y)}{S_1^T(y) dt} S_1^T(y) 
= \lambda_1^T(y) S_1^T(y) dy.
\end{flalign*}
Similarly, we have
	\begin{flalign*}
E  dN(y) (-D)  \widetilde{\omega}(Z,X,y) &= -  \lambda_1^T(y) S_1^T(y); \\
E  dN(y) (1-D) \widetilde{\omega}(Z,X,y) &= -  \lambda_0^T(y) S_0^T(y); \\
   \gamma_1(y) =  E\left\{I(Y\geq y)\widetilde{\omega}(Z,X,y)\right\}     &=  S_1^T(y) - S_0^T(y); \\
	    \gamma_2^m(y) = E\left\{m(D) I(Y\geq y)\widetilde{\omega}(Z,X,y)  \right\}
	   & =  m(1) S_1^T(y) - m(0)S_0^T(y); \\
	    m(1) \gamma_1(y) - \gamma_2^m(y) 
	&=  S_0^T(y) \left\{m(0)-m(1)\right\}; \\
m(0) \gamma_1(y) - \gamma_2^m(y) 
	&=  S_1^T(y) \left\{m(0)-m(1)\right\}; \\
E  dN(y) (-D) \widetilde{\omega}(Z,X,y) &= -  \lambda_1^T(y) S_1^T(y); \\
E  dN(y) (1-D) \widetilde{\omega}(Z,X,y) &= -  \lambda_0^T(y) S_0^T(y).
	\end{flalign*}
Hence 
\begin{flalign*}
RHS = \dfrac{\int -  \lambda_1^T(y) S_1^T(y)   S_0^T(y) \left\{m(0)-m(1)\right\}  dy}{ \int  -  \lambda_0^T(y) S_0^T(y)  S_1^T(y) \left\{m(0)-m(1)\right\} dy } = exp(\psi).
\end{flalign*}

\section{Proof of Proposition \ref{prop:competing}}

Note that
Let $N_d^k(y) = I(Y(d) \leq y, \Delta(d) = 1, \epsilon(d) = k), d=0,1.$
Note that $$ E\left\{ dN^k(y)\dfrac{2Z-1}{f\left( Z|X\right) \delta^D
	\left( X\right) }D\right\}  =  E_X\left[\dfrac{1}{\delta^D \left( X\right) } \sum\limits_{z=0,1}  (2z-1) E\left\{  dN^k(y) D|Z=z,X\right\} 
\right], $$ 
where 
\begin{flalign*}
&	\sum\limits_{z=0,1}  (2z-1) E\left\{  dN^k(y) D|Z=z,X\right\}  \\
&= 	 \sum\limits_{z=0,1}  (2z-1)  E\left\{  dN^k_1(y) D|Z=z,X\right\} \quad \text{(consistency)} \\
&=	 \sum\limits_{z=0,1}  (2z-1) E_{U\mid X} E\left\{ dN^k_1(y)		D|X,U,Z=z\right\}  \quad (Z\ind U\mid X) \\
&= 	 \sum\limits_{z=0,1}  (2z-1)  E_{U\mid X}\left[ E\left\{ dN^k_1(y)|X,U,Z=z\right\} E\left\{ D |X,U,Z=z\right\} \right]   \quad (D\ind T(1), C(1), \epsilon(1) \mid Z, X, U) \\
&= \sum\limits_{z=0,1}  (2z-1)  E_{U\mid X}\left[ E\left\{ dN^k_1(y)|X,U\right\} E\left\{D|X,U,Z=z\right\} \right]  \quad(Z\ind T(1),C(1),\epsilon(1) \mid U,X) \\
&= \delta^D(X) E_{U\mid X}\left[ E\left\{ dN^k_1(y)|X,U\right\}  \right]  \quad (\text{due to   \eqref{eqn:iden}})\\
&= \delta^D(X) E\left\{ dN^k_1(y)|X\right\}
\end{flalign*}
so that
\begin{flalign*}
E\left\{ dN^k(y)\dfrac{2Z-1}{f\left( Z|X\right) \delta^D
	\left( X\right) }D\right\}   &= E\left\{ dN^k_1(y)\right\} 
 \\ &= P(y\leq T(1) < y+dy, C(1)\geq y+dy, \epsilon(1)=k)\\
 &= (dP(T(1) \leq y, \epsilon(1)=k)) P(C(1) \geq y) \\
  &= \dfrac{dP(T(1) \leq y, \epsilon(1)=k))}{S_1^{T,k}(y) dt} S_1^{T}(y) S_1^C(y) \\ &
= \lambda_1^{T,k}(y) S_1^{Y}(y) dy,
\end{flalign*}
where $S_d^Y(y) = P(T(d)\geq y, C(d)\geq y, \epsilon(d) = k) ,d=0,1.$

Similarly,
	\begin{flalign*}
E  dN^k(y) (-D) \omega_0(Z,X) &= -  \lambda_1^{T,k}(y) S_1^Y(y); \\
E  dN^k(y) (1-D) \omega_0(Z,X) &= -  \lambda_0^{T,k}(y) S_0^Y(y).
	\end{flalign*}
	

Hence 
\begin{flalign*}
RHS = \dfrac{\int -  \lambda_1^{T,k}(y) S_1^Y(y)   S_0^{Y}(y) \left\{m(0)-m(1)\right\}  dy}{ \int  -  \lambda_0^{T,k}(y) S_0^Y(y)  S_1^{Y}(y) \left\{m(0)-m(1)\right\} dy } = exp(\psi_k).
\end{flalign*}

\section{Proof of Proposition \ref{prop:ind}}

We shall show that 
$$
    E \left[\int dN(y) e^{-\beta(X) D}  \dfrac{2Z-1}{f(Z\mid X)} \mid X\right] = 0. 
$$
To see this, note that 
\begin{flalign*}
LHS &= E_{U\mid X} E \left[\int dN(y) e^{-\beta(X) D}  \dfrac{2Z-1}{f(Z\mid X)} \mid X,U\right] \\
&= E_{U\mid X} \left\{E[dN(y) e^{-\beta(X) D} \mid Z=1, X,U] - E[dN(y) e^{-\beta(X) D} \mid Z=0, X,U] \right\} \quad (Z\ind U \mid X).
\end{flalign*}

We have
\begin{flalign*}
&\quad E[dN(y) e^{-\beta(X) D} \mid Z, X,U]\\
&= P(D=1\mid Z,X,U) E[ dN_1(y) e^{-\beta(X)} \mid Z,X,U] +  P(D=0\mid Z,X,U) E[ dN_0(y) \mid Z,X,U] \\
&= P(D=1\mid Z,X,U) E[ dN_1(y) e^{-\beta(X)} \mid X,U] +  P(D=0\mid Z,X,U) E[ dN_0(y) \mid X,U] \\
& \quad (Z\ind T(d), C(d) \mid X,U) \\
&= P(D=1\mid Z,X,U) \lambda_1^T(y\mid X,U) S_1^Y(y\mid X,U)e^{-\beta(X)} +  P(D=0\mid Z,X,U)\lambda_0^T(y\mid X,U) S_0^Y(y\mid X,U) \\
&\approx  P(D=1\mid Z,X,U) \lambda_0^T(y\mid X,U) S_1^C(y\mid X,U) +  P(D=0\mid Z,X,U)\lambda_0^T(y\mid X,U) S_0^C(y\mid X,U)  \\
& \quad (T(d) \ind C(d) \mid X,U, S_d^T(y\mid X,U) \approx 1) \\
&=  P(D=1\mid Z,X,U) \lambda_0^T(y\mid X,U) S^C(y\mid X,U,D=1,Z) + \\
&P(D=0\mid Z,X,U)\lambda_0^T(y\mid X,U) S^C(y\mid X,U,D=0,Z)  \quad  (C(d) \ind D,Z \mid X,U) \\
&= P(C\geq y \mid X,U,Z) \lambda_0^T(y\mid X,U) \\
&= P(C\geq y \mid X,U) \lambda_0^T(y\mid X,U)  \quad (Z \ind C \mid X,U)
\end{flalign*}
does not depend on the value $Z$. This finishes the proof.

%

\section{Proof that our data generating mechanism marginalizes to a marginal structural Cox model}

Let $L = (X,U).$
We specify our observed survival model from the following formulation:
\begin{flalign*}
S_{T\mid D,L}(t\mid D=d, L=l) &= \dfrac{\dfrac{S_{T\mid D,L}(t\mid D=d, L=l)}{S_{T\mid D,L}(t\mid D=d, L=l_0)}}{\int \dfrac{S_{T\mid D,L}(t\mid D=d, L=l)}{S_{T\mid D,L}(t\mid D=d, L=l_0)} d F_L(l)} \int S_{T\mid D,L}(t\mid D=d, L=l) dF_L(l)\\
&=  \dfrac{\dfrac{S_{T\mid D,L}(t\mid D=d, L=l)}{S_{T\mid D,L}(t\mid D=d, L=l_0)}}{\int \dfrac{S_{T\mid D,L}(t\mid D=d, L=l)}{S_{T\mid D,L}(t\mid D=d, L=l_0)} d F_L(l)} S_d^T(t),
\end{flalign*}
where the second equality is an application of the g-formula \citep{robins1986new}. 

In our simulation, we let $\dfrac{S_{T\mid D,L}(t\mid D=d, L=l)}{S_{T\mid D,L}(t\mid D=d, L=l_0)} = \exp(\kappa_1 u t + \kappa_2 x t),$ which corresponds to an additive hazards model: 
\begin{flalign*}
S_{T\mid D,L}(t\mid D=d, L=l)&= exp\left(-\int_0^t \Lambda(u) du\right)  \\
&=  exp\left(-\int_0^t (\Lambda_0(u) - \kappa_1 u - \kappa_2 x) du\right)  \\
&= S_{T\mid D,L}(t\mid D=d, L=l_0) exp(\kappa_1 ut + \kappa_2 x t).
\end{flalign*}
The marginal structural Cox model is $S_d^T(t) = (S_0^T(t))^{e^{\psi d}},$ where we let $S_0^T(t) = e^{-\lambda_0 t}.$
Some algebra gives	\begin{flalign*}
S_{T\mid D,L}(t\mid D=d, L=l) &= \dfrac{\exp(\kappa_1 u t + \kappa_2 xt)}{\int \exp(\kappa_1 u t + \kappa_2 xt) dF_L(l)} \exp(-\lambda_0 t e^{\psi d}) \\
&= \dfrac{1}{\lambda_1} (\lambda_1 - \kappa_1 t) \dfrac{1}{\lambda_2} (\lambda_2 - \kappa_2 t)  \exp\left\{(\kappa_1 u + \kappa_2 x - \lambda_0 e^{\psi d}) t \right\}, \numberthis \label{eqn:survival_model}	\end{flalign*}
where the last equality holds since $\kappa_1 < 0, \kappa_2 < 0.$ 

Since the hazard difference is variation dependent of the baseline hazard model, care must be exercised to ensure that \eqref{eqn:survival_model} does give rise to a proper survival model.  First it is obvious that $S_{T\mid D,L}(t = 0\mid D=d, L=l)=1.$ We also have 
$S_{T\mid D,L}(t = \infty\mid D=d, L=l)=0$ since under our parameter specification, $\kappa_1 u + \kappa_2 x - \lambda e^{\psi d} < 0$. Finally, to show that under $\kappa_1=\kappa_2=-1,$
\begin{flalign*}
&\dfrac{d 	S_{T\mid D,L}(t\mid D=d, L=l) }{dt} \\ 
&= \dfrac{1}{\lambda_1\lambda_2}\exp\left\{ -( u +x + \lambda_0 e^{\psi d}) t \right\} \left(  \lambda_1 + t + \lambda_2 + t - (\lambda_1 + t)(\lambda_2+t)  (u + x+ \lambda_0 e^{\psi d})  \right) \leq  0 
\end{flalign*}
we need $\dfrac{1}{\lambda_1+t} + \dfrac{1}{\lambda_2+t} \leq u +x+ \lambda_0 e^{\psi d}.$ Since $t,u$ and $x$ can take any positive values, we need
\begin{flalign*}
\dfrac{1}{\lambda_1} + \dfrac{1}{\lambda_2} &\leq \lambda_0; \\
\dfrac{1}{\lambda_1} + \dfrac{1}{\lambda_2} &\leq \lambda_0 e^{\psi}.
\end{flalign*}
One may verify that our parameter specifications satisfy these conditions.

\section{Additional simulation results}

Table \ref{tab:results5} presented the relative bias of various methods under the same setting as Table \ref{tab:results} in the main file.

\begin{table}[!htbp] 
	\begin{center}
	\small
		\caption{{{Censoring rate and relative bias times 100 (standard error times 100, in parenthesis)  for various methods estimating  the log causal hazard ratio $\psi$.  Here relative bias is defined as the ratio betweeen the bias and the true value of $\psi=0.5$, and ``monotonicity holds'' refers to the case where $\delta^D(X) > 0$ for all $X$. The sample size is 1000. Results are based on 1000 simulated data sets}}}
		\label{tab:results5}
		\begin{tabular}{rccccccccccc}
			& \multicolumn{1}{c}{Censoring \%} &  \multicolumn{6}{c}	{Relative bias$\times 100$  (SE $\times 100$)} \\[5pt]
			\multicolumn{2}{l}{A\ref{assumption:no-interaction} holds ($\beta_3=0$)}	&  Proposed & {\sf Cox-crude} & {\sf Cox-adj} &  {\sf Cox-MSM} & {\sf {\sf MacKenzie}} & {\sf TSPS} & {\sf TSRI} \\[5pt]
			\multicolumn{4}{l}{Monotonicity holds} && \\[2pt]
	$\lambda=0$ &       0 &  0.82(0.44) &  11(0.22) & 3.0(0.23) &  5.0(0.27) &  11(0.35) & $-$7.7(0.35) & 0.52(0.35)   \\
$\lambda=1$ & 17.2\% &  0.95(0.48) &  11(0.24) & 3.1(0.25) &  5.1(0.29) &  12(0.38) & $-$4.8(0.39)  & 0.40(0.39)\\
$\lambda=4$ & 44.9\% &   1.9(0.58) &  11(0.29) & 3.6(0.30) &  5.3(0.34) &  13(0.44) & 0.37(0.47) &  1.3(0.47) \\[8pt]
		\multicolumn{4}{l}{Monotonicity fails} && \\[2pt]
$\lambda=0$ &       0 & $-$0.41(0.45) & 6.2(0.21) & 3.3(0.22) & $-$7.6(0.30) & $-$15(0.53) &  $-$18(0.47) & -6.5(0.49) \\
$\lambda=1$ & 17.0\% & $-$0.75(0.49) & 6.4(0.23) & 3.4(0.23) & $-$8.0(0.32) & $-$15(0.58) &  $-$15(0.53)& -5.9(0.54) \\
$\lambda=4$ & 44.6\% &  $-$1.3(0.60) & 5.7(0.28) & 2.5(0.28) & $-$8.6(0.39) & $-$18(0.69) &  $-$12(0.64) & -7.2(0.65) \\[15pt]
			\multicolumn{2}{l}{A\ref{assumption:no-interaction} fails ($\beta_3=0.5$)}	&   &  &  &   &  &  \\[5pt]
			\multicolumn{4}{l}{Monotonicity holds} && \\[2pt]
$\lambda=0$ &      0 & $-$0.62(0.36) &  10(0.22) & 2.4(0.22) &  4.2(0.30) & 9.1(0.30) &  $-$8.0(0.30) & 0.01(0.29)\\
$\lambda=1$ & 17.2\% & $-$0.58(0.39) &  10(0.24) & 2.6(0.24) &  4.5(0.32) & 9.5(0.33) &  $-$5.4(0.33) & 0.05(0.33) \\
$\lambda=4$ & 44.8\% &  0.11(0.47) &  11(0.28) & 3.3(0.29) &  4.8(0.38) &  11(0.38) & $-$0.45(0.40) &  1.0(0.40) \\[8pt]
		\multicolumn{4}{l}{Monotonicity fails} && \\[2pt]
$\lambda=0$ &       0 &   5.1(0.72) & 7.5(0.20) & 4.2(0.21) & $-$6.5(0.30) & $-$21(0.97) &   $-$16(0.87) & -7.4(0.89) \\
$\lambda=1$ & 17.2\% &   5.0(0.80) & 7.7(0.22) & 4.3(0.22) & $-$6.8(0.32) &  $-$22(1.1) &   $-$14(0.97) & -7.4(0.99) \\
$\lambda=4$ & 44.8\% &   5.1(0.97) & 7.6(0.27) & 4.0(0.27) & $-$6.8(0.40) &  $-$29(1.3) &    $-$14(1.2) &   -10(1.2) \\[5pt]
		\end{tabular}
		\label{result}
	\end{center}
\end{table}

\end{document}